\documentstyle[psfig,nicV]{article}
\begin{document}
%
%Begin Heading
%
\heading{%
Prediction of Thermonuclear Reaction Rates \\
in Astrophysics}
\par\medskip\noindent
%
%Begin Author names
\author{%
Thomas Rauscher
%End Author names
}
%First address
\address{
Departement f\"ur Physik und Astronomie, Universit\"at Basel,
Klingelbergstr.\ 82, CH-4056 Basel, Switzerland
}
\begin{abstract}
Recent
improvements and remaining problems
in the prediction of thermonuclear rates are reviewed.
The main emphasis is on statistical model calculations, but the
challenge to include direct reactions close to the driplines is also
briefly addressed.
Further theoretical as well as experimental investigations are
motivated.
\end{abstract}
\section{Introduction}
The investigation of explosive nuclear burning in astrophysical
environments is a challenge for both theoretical and experimental
nuclear
physicists. Highly unstable nuclei are produced in such processes which
again can be targets for subsequent reactions. Cross sections and
astrophysical reaction rates for a large number of nuclei
are required to perform complete network
calculations which take into account all possible reaction links and do
not postulate a priori simplifications. Most of the involved nuclei are
currently not accessible in the laboratory and therefore theoretical
models have to be invoked in order to predict reaction rates.

In astrophysical applications usually different aspects are emphasized
than in pure nuclear physics investigations. Many of
the latter in this long and well established field were focused on
specific reactions, where all or most "ingredients" (see Sec.~\ref{hf})
were deduced from
experiments. As long as the reaction mechanism is identified properly,
this will produce highly accurate cross sections.
For the majority of nuclei in astrophysical applications such
information is not available. The real challenge is thus not the
application of well-established models, but rather to provide all the
necessary ingredients
in as reliable a way as possible, also for nuclei where no such
information is available. 
%In addition to measurements of relevant cross
%sections, experiments are an indispensable
%means to extract parameter systematics which sometimes still are not
%known very well even along the line of nuclear stability.
%Even with the advent of more and more powerful supercomputers, one has
%to bear in mind that the computational requirement for such
%theoretical approaches should be on a similar scale as, e.g., mass
%models, where the investigation of hundreds and thousands of nuclei is
%possible with manageable computational effort, which is not always the
%case for fully microscopic calculations.

\section{Nuclear Cross Sections and Reaction Rates}

The nuclear cross section $\sigma$ is defined as the number of reactions
$\xi$ target$^{-1}$ s$^{-1}$ divided by the flux $\Phi$ of incoming
particles: $\sigma=\xi / \Phi$. To compute the number of reactions $r$
per volume and time, the velocity (energy) distribution between the
interacting particles has to be considered. Nuclei in an astrophysical
plasma follow a Maxwell-Boltzmann distribution (MBD) and the
thermonuclear reaction rates will have the form \cite{fow67}
\begin{eqnarray}
r_{j,k} & = &\left< \sigma v\right> n_j n_k \nonumber \\
\left< \sigma v\right> :&=&({8 \over {M \pi}})^{1/2} (kT)^{-3/2}
\int_0 ^\infty E \sigma (E) {\rm exp}(-E/kT) dE.
\label{rate}
\end{eqnarray}
Here $M$ denotes the reduced mass of the target-projectile system and
$n_{j,k}$ is the number of projectiles and target nuclei, respectively.
In astrophysical plasmas with high densities and/or low temperatures,
electron screening becomes highly important, which reduces the Coulomb
repulsion.

In the laboratory, the cross section $\sigma^{0\nu}$
for targets in the ground state
is usually measured. 
However, if the plasma is in thermal equilibrium -- and this is
also a prerequisite for the application of the MBD -- the nuclei
will rather be thermally
excited~\cite{arn72}. This has to be accounted for by summing over the excited
target states and weighting each contribution 
with a factor describing
the probability of the thermal excitation.
The ratio of the {\it stellar} cross section $\sigma^*$ and the
laboratory cross section $\sigma^{0\nu}$ is called stellar enhancement
factor (SEF=$\sigma^*/\sigma^{0\nu}$).
The stellar reaction rate is then obtained by inserting $\sigma^*$ into
Eq.~(\ref{rate}): $r^*=\left< \sigma^* v \right> n_j n_k = \left<
\sigma v \right>^* n_j n_k$.
It should be noted that only the stellar rate $r^*$ (involving the
stellar cross section) always obeys
reciprocity (because $\sigma^*$ does) and that
therefore only $r^*$ can be used to compute the reverse rate.
Thus, it is very important when measuring reaction cross sections in the
laboratory for astrophysical application to measure the cross section
for the reaction in the direction that is least affected by excited states in
the target. This is almost always the exoergic reaction (i.e.\ $Q>0$). Even
so, the stellar rate at temperatures in excess of a few billion degrees will
vary considerably from that of a determination based on targets in their
ground states. The experimentally determined laboratory rate should then
be multiplied by the
SEF, which can, in most cases, only
be determined by a theoretical calculation.
Even at the low temperatures of the $s$ process, the SEF can already be
important. Neutron capture cross 
sections of isotopes in the rare earth region have recently been measured 
to such a high
accuracy that the knowledge of precise SEFs becomes crucial for further
constraining $s$ process conditions~\cite{voss98}. In principle,
sufficiently reliable SEF calculations would provide an additional
thermometer for the $s$-process environment, completely independent of
the usual estimates via $s$-process branchings \cite{KBW89}.

In general, the cross section will be the sum of
the cross sections resulting from compound reactions
via an average over overlapping resonances (HF) and via single resonances
(BW), direct reactions (DI) and interference terms:
\begin{equation}
\label{sumcs}
\sigma(E)=\sigma^{\mathrm{HF}}(E)+
\sigma^{\mathrm{BW}}(E)+
\sigma^{\mathrm{DC}}(E)+
\sigma^{\mathrm{int}}(E) \quad .
\end{equation}
Depending on the number of levels per energy interval in the system
projectile+target, different reaction mechanisms will dominate 
\cite{wag69,rau97}.
Since different regimes of level densities are probed at the various
projectile energies, the application of a specific description depends
on the energy. In astrophysics, one is interested in energies in the
range from a few tens of MeV down to keV or even thermal energies
(depending on the charge of the projectile). For instance, when varying
the energy of a neutron beam from 10 MeV down to thermal energy, the
resulting cross sections will be dominated by mainly three different
contributions: At the highest energy, many close resonances overlap and allow
to use an average cross section calculated in the statistical model
(Hauser-Feshbach, HF). To lower energies, the nuclear states become more
and more widely spaced until one can identify single resonances which
can be included into the HF equation as a special case,
yielding the well-known Breit-Wigner (BW) shape.
In between resonances the
cross sections are determined by the tail of resonances and the direct
(DI) contribution. At the lowest energies, the levels can be so
widely spaced that the cross section is described well by the direct
component alone. Extrapolations from one regime into another can be
extremely misleading.

The relevant nuclear levels are found in the effective energy window of
a reaction, i.e.\ the energy range which is mainly contributing to the
determination of the nuclear reaction rate. This energy window is
usually well defined, due to the sharply peaking integrand of
Eq.~(\ref{rate}). For neutrons, it is given by the width of the peak of
the MBD. For charged
particles the cross section includes the penetrability through the
Coulomb barrier which is exponentially increasing with increasing
energy. Folding this cross section with the velocity distribution gives
rise to a broader peak shifted to higher energies, the so-called Gamow
peak. Using experimental information or a theoretical level density
description, it is possible to determine the number of levels within the
effective energy window and thus derive the applicability of the
statistical model as a function of temperature~\cite{rau97}. Below a
critical temperature, averaging over too few resonances is not
appropriate anymore and the HF will misjudge
the cross section. The critical temperature is especially high for
targets with closed shells which exhibit widely spaced nuclear levels,
and for targets close to the driplines which have low particle
separation energies (and $Q$ values).

\section{The Statistical Model}
\label{hf}

The majority of nuclear reactions in astrophysics
can be described in the framework of the
statistical model
(HF)~\cite{hau52}. This description assumes that the reaction
proceeds via a compound nucleus which finally decays into the reaction
products. With a sufficiently high level density, average cross sections
\begin{equation}
\sigma^{\rm HF}=\sigma_{\rm form} b_{\rm dec} = \sigma_{\rm form} 
{\Gamma_{\rm final} \over \Gamma_{\rm tot}}
\end{equation}
can be calculated which can be factorized into a cross section
$\sigma_{\rm form}$ for the formation of the compound nucleus and a
branching ratio $b_{\rm dec}$, describing the probability of the decay
into the channel of interest compared with the total decay probability
into all possible exit channels. The partial widths $\Gamma$ as well as
$\sigma_{\rm form}$ are related to (averaged) transmission coefficients, 
which comprise the central quantities in any HF calculation.

Many nuclear properties enter the computation of the transmission coefficients:
mass differences (separation energies), optical potentials, GDR widths,
level densities. The transmission coefficients can be modified due to
pre-equilibrium effects which are included in width fluctuation
corrections~\cite{tep74} (see also \cite{rau97} and references therein)
and by isospin effects. It is in the description of the nuclear
properties where the various HF models differ.

In the following sections, the most important ingredients and the
usual pa\-ra\-me\-tri\-za\-tions used in astrophysical applications are briefly 
discussed. A choice of what is thought of being
the currently best parametrizations is
incorporated in the new HF code NON-SMOKER~\cite{nonsmoker}, which is
based on the well-known code SMOKER~\cite{thi87}.

\subsection{Optical Potentials}

Early astrophysical studies (e.g. \cite{arn72,hol76,woo78}) made use of
simplified equivalent square well potentials and the black nucleus
approximation. It is equivalent to a fully absorptive potential, once a
particle has entered
the potential well and therefore does not permit resonance effects.
This leads to deviations from experimental data at low energies,
especially in mass regions where broad resonances in the continuum can
be populated~\cite{hof98}. An additional effect, which is only
pronounced for $\alpha$ particles, is that absorption in the Coulomb
barrier~\cite{mic70} is neglected in this approach.

Improved calculations have to employ appropriate {\it global} optical
potentials which make use of imaginary parts describing the absorption.
The situation is different for nucleon-nucleus and
$\alpha$-nucleus potentials. Global optical potentials are quite well
defined for neutron and protons. It was shown~\cite{thi83,cow91}
that the best fit
of s-wave neutron strength functions is obtained with the optical
potential by~\cite{jeu77}, based on microscopic
infinite nuclear matter calculations for a given density,
applied with a local density approximation.
It includes corrections of the imaginary part~\cite{fantoni81,mahaux82}.
A similar description can be used for protons. Numerous experimental
data document the reliability of the neutron potential for astrophysical
applications. For protons, data are more scarce but recent investigations
\cite{bork98} also show good agreement.

In the case of $\alpha$-nucleus potentials, there are only few {\it
global} parametrizations available at astrophysical energies.
Most global potentials are of the Saxon--Woods form,
parametrized at energies above about 70 MeV,
e.g.~\cite{sin76,nol87}.
The high Coloumb barrier makes a
direct experimental approach very difficult at low energies. More
recently, there were attempts to extend those parametrizations to
energies below 70 MeV~\cite{avr94}. Astrophysical calculations mostly
employed a phenomenological Saxon--Woods
potential based on extensive data~\cite{mcf66}.
This potential is an energy-- and mass--independent mean
potential. However, especially at low energies the imaginary part of the
potential should be highly energy--dependent. Nevertheless, this
potential proves to be very successful in describing HF cross sections.
It failed so far only in one case, the recently measured
$^{144}$Sm($\alpha$,$\gamma$)$^{148}$Gd low-energy cross section~\cite{som98}.
Nevertheless, this showed that future improved $\alpha$ potentials have
to take into account the mass- and energy-dependence of the potential.
Several attempts have been made to construct such an improved potential.
Extended investigations of $\alpha$ scattering data~\cite{mohr94,atz96}
have shown that the data can best be described with folding
potentials~\cite{sat79}. They also found a systematic mass- and
energy-dependence.
Very recently,
that description was used for a global
approach~\cite{nonsmoker,hirschegg}. The idea is to parametrize the data
including nuclear structure information.
The accuracy reached~\cite{hirschegg} is comparable to the potential of
Ref.~\cite{mcf66}. The same approach was used by \cite{gra98}, without
including further microscopic information.
However, the limitation of this method is that a Woods-Saxon term with
fixed geometry is still used for the imaginary part. The resulting
transmission coefficients are very sensitive to the shape of the
imaginary part, which leads to an ambiguity in the parametrization.
Experimental data indicate that the geometry may also be
energy-dependent~\cite{avr94,mohr97,bud78}. This can be
understood in terms of the semi-classical theory of elastic
scattering~\cite{bri77} which shows that with varying energy
different radial parts of the potential are probed. The effect can be
explicitly considered in a global potential~\cite{ringberg}.
Nevertheless, more experimental data are needed which should be
consistently analyzed at different energies.
Further complications arise from the fact that it is yet unclear if
potentials extracted from scattering data can indeed describe
transmission coefficients well~\cite{avr94}. Clearly, further effort has
to be put into the improvement of global $\alpha$-nucleus potentials at
astrophysically relevant energies.

\subsection{$\gamma$ Width}

The $\gamma$-transmission coefficients have to include the dominant E1
and M1 $\gamma$ transitions. The smaller, less important M1 transitions
have usually been treated with the simple single particle approach
$T\propto E^3$~\cite{bla52}. The E1 transitions are usually calculated
on the basis of the Lorentzian representation of the Giant Dipole
Resonance (GDR). Many microscopic and macroscopic models have been
devoted to the calculation of GDR energies and widths. Analytical fits
as a function of mass number $A$ and charge $Z$ were also used in
astrophysical calculations~\cite{hol76,woo78}. An excellent fit to the
GDR energies is obtained with the hydrodynamic droplet
model~\cite{mye77}. An improved microscopic-macroscopic
approach is used in most modern
reaction rate calculations, based on dissipation and the coupling to
quadrupole surface vibrations (see~\cite{cow91}).

Most recently it was shown~\cite{gor98} that the inclusion of pygmy
resonances might have important consequences on the E1 transitions in
neutron-rich nuclei far off stability. The pygmy resonances can be caused
by a neutron skin which generates soft vibrational modes~\cite{isa92}.
While the effect close to stability is small, neutron capture cross
sections could be considerably enhanced close to the neutron dripline.

\subsection{Level Density}

Until recently, the nuclear level density (NLD) has given rise to the largest
uncertainties in the description of nuclear
reactions~\cite{woo78,cow91}. 
For large scale astrophysical applications
it is necessary to not only find reliable methods for NLD
predictions but also computationally feasible ones. 
Such a model is the
non-interacting Fermi-gas model. Most statistical model calculations use
the back-shifted Fermi-gas description~\cite{gil65}. More sophisticated
Monte Carlo shell model calculations (e.g.~\cite{dea95}), as well as
combinatorial approaches (e.g.~\cite{paa97}), have shown excellent
agreement with this phenomenological approach and justified the
application of the Fermi-gas description.
While different fits to different mass regions to
obtain the free parameters were performed in many
investigations~\cite{hol76,woo78,cow91}, a most recent study was able to
arrive at considerably improved NLDs with fewer parameters in
the mass range $20\leq A\leq 245$~\cite{rau97}. They applied an
energy-dependent NLD parameter~\cite{ign} together with
microscopic corrections from nuclear mass models. 
The fit to
experimental NLDs is also better than a recent analytical BCS
approach~\cite{gor96} which implemented level spacings from a
microscopic mass model. (In fact, see~\cite{pea96,doba96} for doubts
on the reliability of the BCS model for neutron-rich nuclei).
%Another approach adopted an analytical BCS model implementing level
%spacings from a microscopic mass model~\cite{gor96}. This lead to
%inferior predictions at the line of stability as compared to the
%Fermi-gas approach but a more reliable extrapolation far from stability
%was claimed. However, there are also claims that the BCS model may
%become unreliable for neutron-rich nuclei~\cite{pea96,doba96}.

Further work has to be invested in the problem of the prediction of
the parity distribution at low excitation energies of the nucleus.

\subsection{Isospin Effects}

The original HF equation~\cite{hau52} implicitly assumes complete
isospin mixing but can be generalized to explicitly treat the
contributions of the dense background states with isospin $T^<=T^{\rm
g.s.}$ and the isobaric analog states with
$T^>=T^<+1$~\cite{gri71,har77}. The inclusion of the isospin treatment
has two major effects
on statistical cross section calculations in astrophysics~\cite{nonsmoker}:
the suppression of $\gamma$ widths for reactions involving
self-conjugate nuclei and the suppression of the neutron emission
in proton-induced reactions. (Non-statistical effects, i.e.\ the
appearance of isobaric analog resonances, will not be discussed here.)
Firstly, in the case of ($\alpha$,$\gamma$) reactions on targets with $N=Z$, the
cross sections will be heavily suppressed because $T=1$ states cannot be
populated due to isospin conservation. A suppression will also be found
for capture reactions leading into self-conjugate nuclei, although
somewhat less pronounced because $T=1$ states can be populated according
to the isospin coupling coefficients.
In previous reaction rate calculations~\cite{woo78,cow91} the
suppression of the $\gamma$--widths was treated completely
phenomenologically by employing arbitrary and mass-independent
suppression factors.
In the new NON-SMOKER code~\cite{nonsmoker}, the
appropriate $\gamma$ widths are automatically obtained, by explicitly 
accounting for $T^<$ and $T^>$ states.

Secondly, assuming incomplete isospin mixing, the strength of the
neutron channel will be suppressed in comparison to the proton channel in
reactions p+target~\cite{gri71,sar82}. This leads to a smaller
cross section for (p,n) reactions and an increase in the cross section
of (p,$\gamma$) reactions above the neutron threshold.
Such an effect has recently been found
in a comparison of experimental data and NON-SMOKER
results~\cite{bork98}.

\section{Direct Reactions}

As stated above, the HF approach can only be applied for
sufficiently high NLDs~\cite{rau97}. At low NLDs, 
the other
terms in Eq.~(\ref{sumcs}) will begin to dominate. Many investigations
(e.g.~\cite{ohu96,mei96,bee95,kra96,mohr98,mohr98a}) have been devoted 
to the calculation of direct neutron 
capture for light nuclei and nuclei close to magic neutron numbers. 
Utilizing folding
potentials, these calculations can yield reliable cross sections
provided that information on the bound states and the spectroscopic
factors is known.
Even in the regime of single resonances, the feeble DI contribution can
be seen nowadays, when comparing highly precise resonance data and 
activation measurements (e.g.~\cite{koe98}).

The prediction of the DI contribution to neutron capture cross sections
close to the dripline (which may be important in the $r$ process)
remains a challenge. Far off stability, the required nuclear properties
are not known and have to be taken from other theories~\cite{mat,gor97,rau98}. 
However, it was shown~\cite{rau98} that a straightforward application
produces cross sections which are highly sensitive to slight changes in
the predicted masses and level energies. Furthermore, it is not yet
clear which spectroscopic factors to employ and how to model
interference between DI and HF in a simple manner. Further work is clearly
needed.

\section{Summary}

The new generation of HF models can make reliable predictions of nuclear
cross sections. Furthermore, the applicability range of HF has been
quantified and thus the boundary between different reaction mechanisms
clarified. Although the phenomenological parametrizations of nuclear
properties already display good quality, there is a clear need for more
experimental data for checking and further improving current models.
Especially investigations over a large mass range would
prove useful to fill in gaps in the knowledge of the nuclear structure
of many isotopes and to construct more powerful parameter systematics,
which sometimes are badly known even at the line of stability.
Such investigations should include neutron-,
proton- and $\alpha$-strength functions, as well as radiative widths,
and charged particle scattering and reaction cross sections for {\it stable}
and unstable isotopes. More capture data with
self-conjugate final nuclei would also be highly desireable.

The new code NON-SMOKER~\cite{nonsmoker}
contains updated nuclear information as well as additional
effects. The NON-SMOKER reaction rate library is electronically available at
{\it http://quasar.physik.unibas.ch/\~{ }tommy/reaclib.html} .

\acknowledgements{This work was partially supported by the Swiss
Nationalfonds (grant 20-47252.96). TR is an APART fellow of the Austrian
Academy of Sciences.}

\vfill
\end{document}